\documentclass[3p,times]{elsarticle}

\usepackage{ecrc}

\usepackage{epstopdf}
\usepackage{graphicx}
\usepackage{subfigure}
\usepackage{amsmath,amssymb}
\volume{00}

\firstpage{1}

\journalname{Nuclear Physics A}

\runauth{H.-T. Ding}

\jid{nupha}

\jnltitlelogo{Nuclear Physics A}

\newcommand{\nc}[1]{\newcommand{#1}}
\nc{\beqa}{\begin{eqnarray}}
\nc{\eeqa}{\end{eqnarray}}

\nc{\beq}{\begin{equation}}
\nc{\eeq}{\end{equation}}
\nc{\hmu}{\hat{\mu}}
\nc{\nn}{\nonumber}

\nc{\vecp}{\mathbf{p}}
\nc{\md}{\mathrm{d}}
\nc{\vecx}{\mathbf {x}}
\nc{\vecnull}{\mathbf {0}}

\begin{document}

\begin{frontmatter}



%% e.g. \dochead{17th International Conference on Dynamical Processes in Excited States of Solids}

\title{Recent lattice QCD results and phase diagram of strongly interacting matter}

%% \author[label1]{}

\author{Heng-Tong Ding}

\address{Key Laboratory of Quark \& Lepton Physics (MOE) and Institute of
Particle Physics, \\
Central China Normal University, Wuhan 430079, China}

\begin{abstract}

I  review recent lattice QCD results on a few selected topics which are 
relevant to the heavy ion physics
community. Special emphasis is put on the QCD equation of state at vanishing and nonzero baryon chemical potential,
the onset of deconfinement of open strange and charmed hadrons, the
contribution from experimentally
not yet observed hadrons to the thermodynamics of the hadronic gas and 
their influence on freeze-out conditions of strange and light-quark hadrons.
\end{abstract}


\end{frontmatter}

% \linenumbers

\section{Introduction}

Lattice QCD is a discretized version of QCD in the Euclidean space time which reproduces QCD when the lattice spacing goes to
zero, that is in the continuum limit. Most lattice QCD calculations which are 
at present available for use by the heavy ion community have been
performed using non-chiral fermions which recover the flavor or chiral symmetry of QCD only in the continuum limit, e.g mostly used are staggered and 
Wilson fermions. Chiral fermions are generally much more expensive to work with. However, thanks to Moore's law, currently these actions are also used and
start to produce interesting results on QCD thermodynamics, e.g. the 
value of the crossover temperature $T_c$ has been confirmed~\cite{DWF_Tc}
and investigations of the restoration of $U(1)_A$ symmetry are also going 
on~\cite{axial}.

\section{Equation of state at vanishing and non-zero baryon density}
\label{intro}

\begin{figure}[!th]
\begin{center} 
\includegraphics[width=0.32\textwidth]{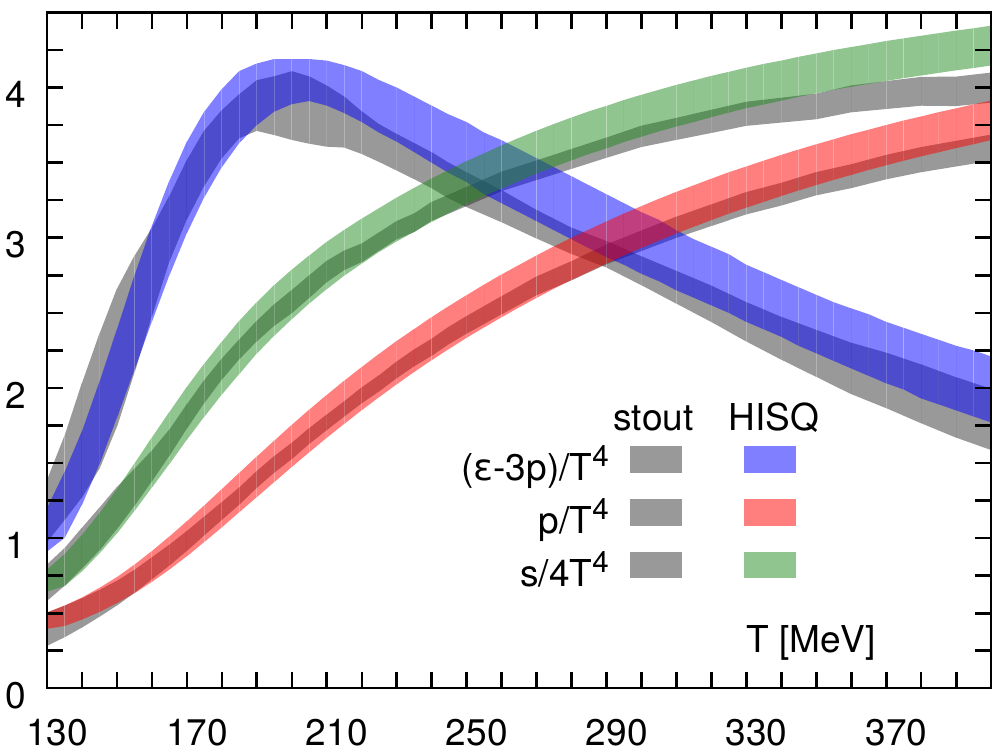}~~\includegraphics[width=0.32\textwidth]{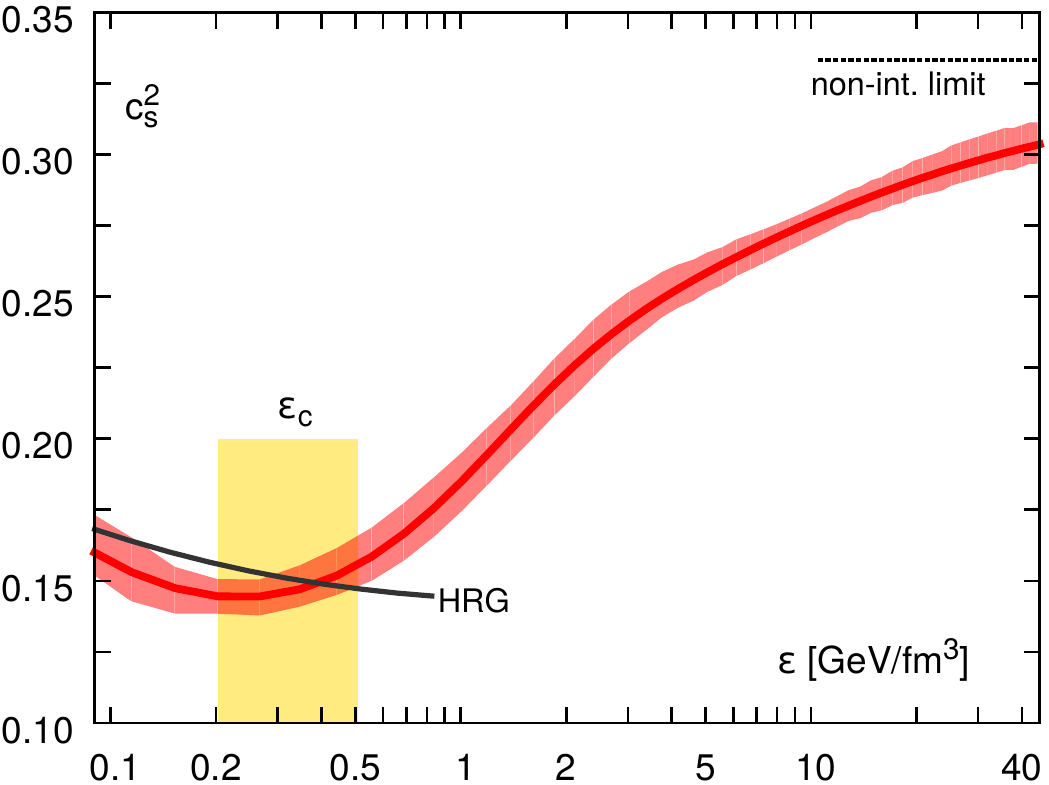}
\caption{Left: Comparisons of continuum extrapolated results of  trace anomaly $(\epsilon - 3p)/T^4$, pressure $p/T^4$ and entropy density $s/T^4$
 from Wuppertal-Budapest (stout)~\cite{EOS_WB} and HotQCD (HISQ) collaborations~\cite{EOS_hotQCD}. Right: The speed of sound squared from lattice QCD calculations using the HSIQ action
 and the HRG model as a function of energy density. Figures are
taken from Ref.~\cite{EOS_hotQCD}.}
\label{fig:EOS}
\end{center}
\end{figure}

The equation of state (EoS) of QCD matter contains information on the change 
of degrees of freedom 
in different temperature as well as different baryon density regimes and reflects the transition between different states 
of the QCD matter. It is one of the important ingredients to understand the evolution of the fireball produced in
heavy ion collisions through classical hydrodynamic equations. The computation 
of the QCD EoS has been one of the major goals 
in the lattice QCD community since 1980~\cite{firstlQCDEoS}.

Most of current EoS calculations are performed using the so-called staggered 
action. The shortfall of this action is that it breaks flavor symmetry
at finite lattice spacing. The consequence of this is that in the staggered 
formulation of lattice QCD there are additional 15 heavier unphysical pions 
in addition to 1 physical Goldstone pion which lead to the physical pion 
spectrum only in the continuum limit. There are ways to improve this 
action at finite values of the lattice cutoff. One of them is the stout 
action, mostly used by Wuppertal-Budapest collaboration;  another one is the 
Highly Improved Staggered Quark (HISQ) action which is mostly used by the 
HotQCD collaboration.
At zero temperature the stout as well as the HISQ actions greatly reduce the 
masses of these 15 unwanted pions. Moreover,  in the infinite temperature 
limit quantities calculated with the HISQ action, e.g. the pressure and 
energy density of the
quark gluon plasma, approach their Stefan-Boltzmann limits faster than in calculations
performed with the stout action at the same value of the lattice cutoff. 
However, continuum extrapolations of 
observables using either of these two actions can be safely performed on 
lattices at moderate values of the lattice cutoff.

In this conference the HotQCD collaboration has presented their recent results on the QCD EoS at vanishing baryon density~\cite{EOS_Bazavov}.
Data with $N_t=4$, 6, 8, 10 and 12 have been computed and 
$N_t$=8, 10 and 12 have been used 
for the continuum extrapolation. Comparisons between results from the
Wuppertal-Budapest (stout) and HotQCD (HISQ) collaborations are shown in the 
left panel of Fig.~\ref{fig:EOS}. It can be clearly seen that after the continuum
extrapolations quantities obtained from stout and HISQ actions, i.e. trace anomaly $(\epsilon - 3p)/T^4$, pressure $p/T^4$ and entropy density $s/T^4$, 
are in good agreement in the temperature range from 130 MeV to 400 MeV. A 
noticeable difference shown in the entropy density reaches about 7\% at a
temperature of 400 MeV. The resolution of this discrepancy requires more 
detailed calculations of the trace anomaly at higher temperatures.
The right panel in Fig.~\ref{fig:EOS} shows the speed of sound squared from lattice QCD calculations and the HRG model as a function of energy density.
The chiral crossover temperature region, i.e. $T_c\in$ [149,163] MeV, is 
indicated by the yellow band in the plot. Here the value of the critical 
energy density $\epsilon_c$ lies in
the range of (180 - 500) MeV/fm$^3$. At the softest point of the speed of 
sound squared the energy density is only slightly larger than that of normal 
nuclear matter, i.e.
$\epsilon_{nuclear}$=150 MeV/fm$^{3}$. One may compare this energy density also
to that 
inside a proton, $\epsilon_{proton}$= 450 MeV/fm$^3$, and the energy density 
of an ideal quark gluon plasma in the crossover over temperature region, which 
ranges from 900 MeV/fm$^3$ to 1440 MeV/fm$^3$.

\begin{figure}[!th]
\begin{center} 
\includegraphics[width=0.36\textwidth]{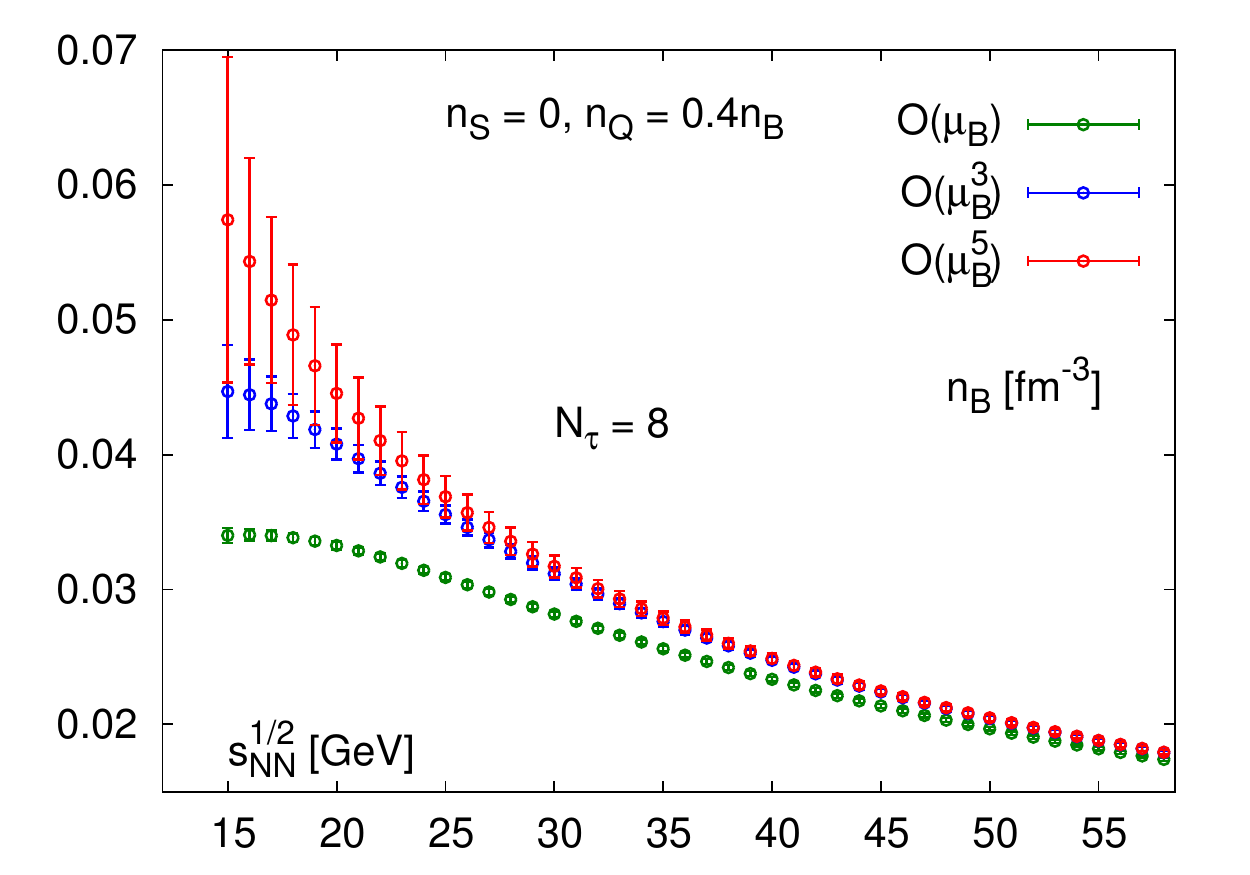}~~\includegraphics[width=0.43\textwidth]{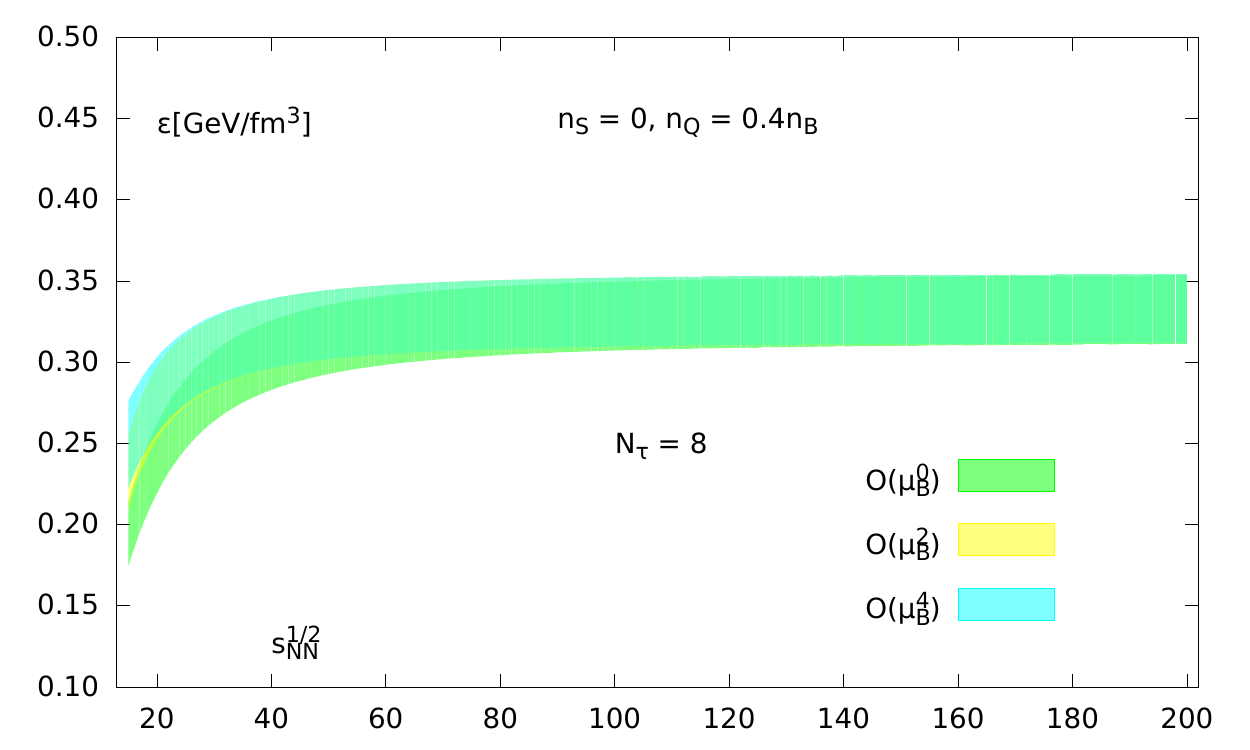}
\caption{Baryon number density $n_B$ (left) and energy density $\epsilon$ (right) as function of $\sqrt{s_{NN}}$ along the freeze-out line. Calculations have been performed 
using the Taylor expansion method upto the order of $\mu_B^5$ and $\mu_B^{4}$, respectively. The freeze-out curve
is obtained using the parameterization in Ref.~\cite{Cleymans:1999st}. Figures are taken from Ref.~\cite{EOS_fmu}.}
\label{fig:EOSmu}
\end{center}
\end{figure}

First results for the equation of state at nonzero baryon density obtained
with the HISQ action in a Taylor series upto  
4h order have been reported by the BNL-Bielefeld-CCNU collaboration in this conference~\cite{EOS_fmu}. These calculations are performed on lattices with $N_\tau=6$ and 8. The 2nd order results from this study are consistent with those using the stout action~\cite{EOS_fmuWB}. As seen from the left panel of Fig.~\ref{fig:EOSmu} a reliable description of baryon number density $n_B$ down to $\sqrt{s_{NN}}\sim 30$~GeV can be obtained by including terms up to $\mathcal{O}(\mu_B^3)$; corrections from the next higher order, i.e. $\mathcal{O}(\mu_B^5)$, reach about 30\% at $\sqrt{s_{NN}}\sim 15$~GeV. It is interesting to see in the right panel of Fig.~\ref{fig:EOSmu} that the energy density
$\epsilon$ stays almost constant, $\epsilon \sim 0.33$~GeV/fm$^3$, down to 
$\sqrt{s_{NN}}\sim 50$~GeV. Note that at zero baryon chemical potential in the chiral crossover temperature region the critical energy density is
$\epsilon_c\in$[0.18, 0.5] GeV/fm$^3$.

\section{Deconfinement aspects of the QCD transition}

It is now well established that the QCD transition is just a rapid crossover 
at a transition temperature  $T_c\sim$155 MeV~\cite{WB_Tc,HotQCD_Tc}. 
Its chiral aspects manifest themselves 
in chiral quantities, such as the light quark number susceptibilities or 
the light quark chiral condensates, while the deconfinement aspects are related 
to the liberation of light degrees of freedom seen in the rapid change of
the energy density or the change of
heavier-quark degrees of freedom. As proposed long time ago by Matsui and Satz~\cite{HQ_seed} quarkonium states can survive at temperatures above 
the critical temperature $T_c$. The fate of heavy quarkonium states has been studied extensively recently on the lattice~\cite{Ding:2014xha}. Currently there 
are at 
least three approaches actively pursued to detect the modification of heavy 
quark bound states on the lattice. 
The first approach is to calculate the screening masses of a heavy quark-antiquark pair
from spatial correlation functions. 
By comparing the screening masses at different values of the temperature with 
corresponding
pole masses, i.e. the mass of quarkonium states in the vacuum, one can get some 
information on thermal modifications of bound states. The calculation of 
screening masses is straightforward and not expensive. However, the screening 
mass itself contains no more information other than the thermal modification 
of the pole mass~\cite{Mscr}. Establishing a link to the melting of states is
ambiguous. The second approach aims at the determination of the heavy 
quark potential from which spectral functions of quarkonium states can be 
obtained by using the Schr\"odinger equation. The imaginary and real parts of
the heavy quark potential in general are related to the peak location and 
width of the spectral function of 
Wilson line correlation functions~\cite{Burnier:2012az}. 
This approach is hampered by the noise to signal ratio of the Wilson line correlators which makes a
precise determination of the width of the spectral function difficult. 
Recent progress has been made on developing an improved method 
to extract the spectral function from Wilson line correlators~\cite{Burnier:2013nla}. 
A third approach is to extract meson spectral functions from temporal 
Euclidean correlation functions. 
As the value of heavy quark mass $M$ is large the lattice spacing has to be small 
enough, i.e. $Ma \ll 1$ to accommodate the heavy quark on the lattice. 
One way out is to directly calculate the heavy quarkonia correlation function by brute force, 
i.e. by reducing the lattice spacing to a sufficiently small value. 
Another way is to make use of effective theories, e.g. non-relativistic QCD, 
which however is appropriate only for bottom quarks and unfortunately does 
not allow to perform a continuum limit. 
Current state-of-the-art calculations, performed within
quenched lattice QCD, suggests that all the charmonia states are melted at 
$T\geq 1.5~T_c$~\cite{charm}.  Recent exploratory studies that also include 
dynamic quark degrees of freedom give consistent results~\cite{charm2}. 
For bottomonia it has recently been found that $S$ wave states start to melt at a temperature larger than 2$~T_c$ while $P$ wave states melt already 
immediately above $T_c$~\cite{Bottomonia}.

The fate of heavy-light mesons or baryons also reflects the change of the 
relevant degrees of freedom in strong interaction matter. 
For instance, the abundance of strange hadrons is considered
as one of the signals that Quark Gluon Plasma is formed. In the heavy-light systems the net quantum number carried by the heavier quark, e.g. strangeness
or charm,
is nonzero rather than zero in the case of  heavy quark-antiquark systems. 
As the electric charge 
$Q$ and baryon number $B$ of hadrons are integer quantum numbers in the 
confined hadronic phase but are fractional numbers in the 
deconfined QGP phase,
fluctuations and correlations of these quantum numbers with strangeness
or charm allow to probe the deconfinement of carriers of strangeness and 
charm degrees of freedom, i.e. the strange and charm quarks.
These fluctuations are defined as the derivatives of pressure with respect to 
the chemical potential of a given quark flavor. In practice one needs 
to construct certain ``order parameters" that vanish (approximately) in one 
phase and are large and nonzero in the other phase.

\begin{figure}[!th]
\begin{center} 
\includegraphics[width=0.35\textwidth]{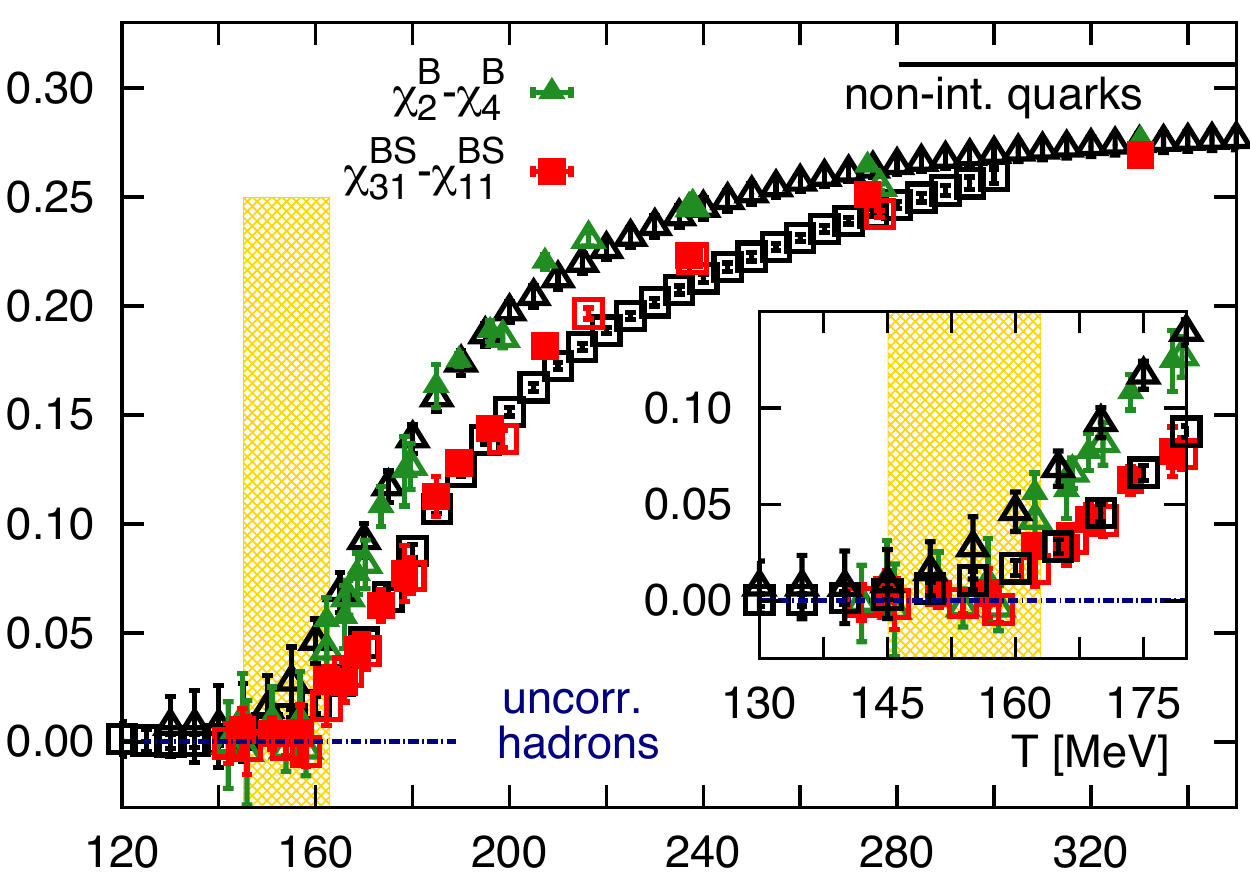}\includegraphics[width=0.35\textwidth]{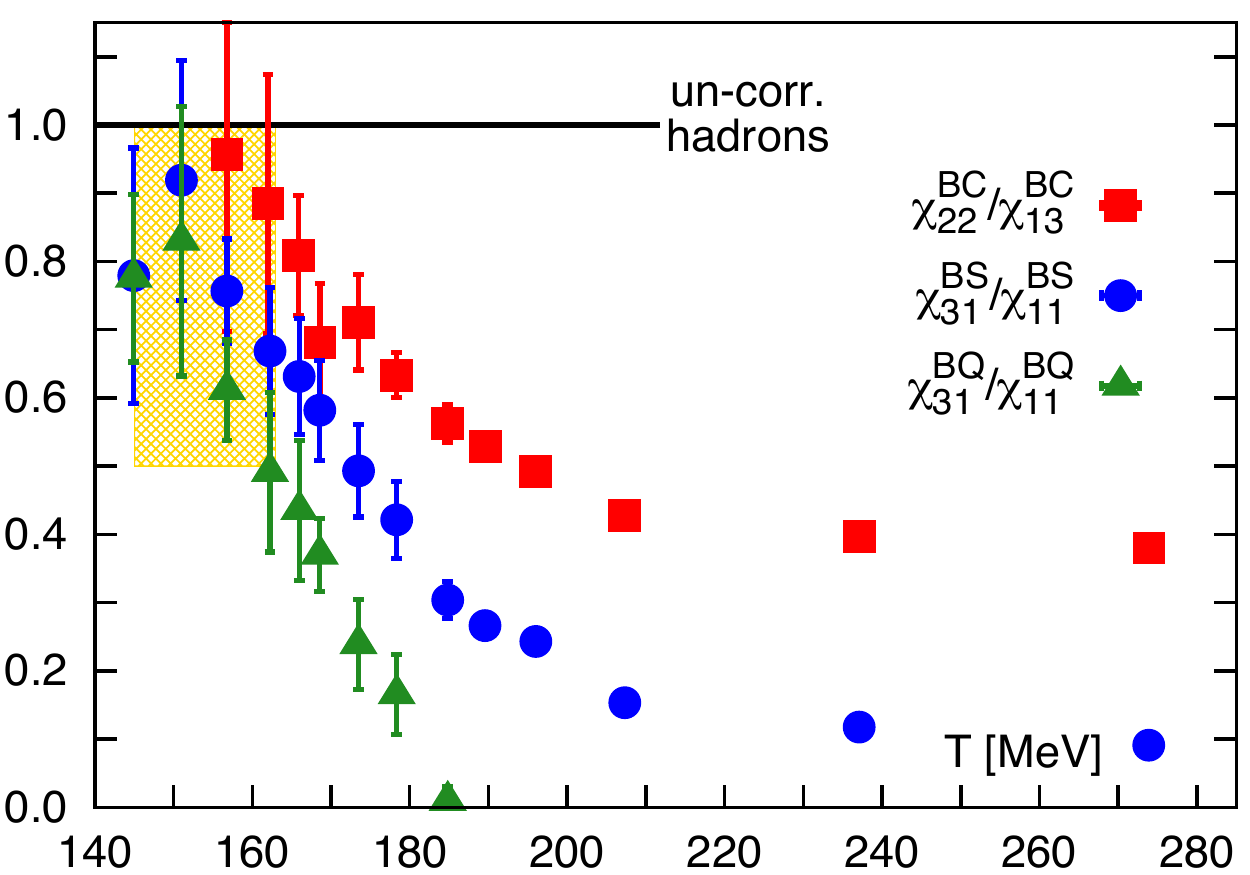}
\caption{Left: Onset of the dissociation of open strange hadrons in the chiral crossover temperature region. Black points are results obtained using the stout
action~\cite{Bellwied:2013cta} while the other data points are those obtained using the HISQ action~\cite{strangeness}. Right: Onset of open charm and open strange hadrons 
happen in the chiral crossover temperature region as seen from ratios of baryon-charm (BC),
baryon-strangeness (BS) and baryon-electric charge (BQ) correlations~\cite{charmness}. Yellow bands shown in these two plots represent the temperature 
window of the chiral crossover $T_c=154\pm 9$ MeV.}
\label{fig:decon}
\end{center}
\end{figure}

The starting point for such an analysis is the description of strangeness in 
an uncorrelated hadron gas. 
In the Hadron Resonance Gas (HRG) model, due to their large masses, heavy mesons 
and baryons follow Boltzmann statistics. The pressure of all the strange hadrons in an uncorrelated hadron resonance gas, $P^{HRG}_S$, can be decomposed
into a mesonic, $P^{HRG}_M$, and baryonic part, $P^{HRG}_B$~\cite{strangeness}
\beq
P^{HRG}_S(\mu_B,\hmu_S) = P^{HRG}_{|S|=1,M} \cosh(\hmu_S)  + \sum_{\ell=1}^{3}P^{HRG}_{|S|=i,B} \cosh(\hmu_B-\ell\hmu_S)\; .
\label{eq:P-HRG}
\eeq
Fluctuations of conserved quantum numbers are obtained as derivatives of the
pressure with respect to various chemical potentials 
$\hat{\mu}_X=\mu_X/T$, i.e.
$\chi_{\rm mn}^{\rm XY} =\frac{\partial^{(m+n)} \big{(}P(\hat{\mu}_X,\hat{\mu}_Y)/T^4\big)}{\partial \hat{\mu}_X^m \partial \hat{\mu}_Y^n}\Big{|}_{\vec{\mu}=0}$
where $X,Y=B,Q,S,C$ and $\vec{\mu}=(\mu_B,\mu_Q,\mu_S,\mu_C)$ and $\chi_{0n}^{XY}\equiv\chi_n^Y$ and $\chi_{m0}^{XY}\equiv\chi_m^X$. 
Baryon-strangeness correlations that differ by an even 
number of derivatives with respect to $\mu_B$, are identical in a HRG. This
is easily seen from the above equation. Combinations of conserved charge
correlations such as $\chi^{\rm BS}_{31} - \chi^{BS}_{11}$ or 
$\chi_2^B-\chi_4^B$ thus vanish in an uncorrelated gas of hadrons within the classical Boltzmann approximation.
Here $\chi^{\rm BS}_{31} - \chi^{BS}_{11}$ receive contributions only from strange hadrons while $\chi_2^B-\chi_4^B$ receive contributions from all hadrons. It is apparent from the left panel in Fig.~\ref{fig:decon} 
that all three quantities shown there would vanish in a HRG. They deviate 
from zero at almost the same temperature, i.e. in the vicinity of
the chiral crossover temperature as indicated by the yellow band. 
This suggests that strange quark degrees of freedom start to get liberated 
from strange hadrons at almost the same temperature as light quarks.
The whole analysis has been repeated using a different fermion action in 
Ref.~\cite{Bellwied:2013cta}.

The same concepts can also be applied to the open charm hadrons~\cite{charmness}. Since the charmed baryon sector with charm quantum number $C=\pm 1$ dominates the contribution 
to the partial pressure of all charmed baryons in an uncorrelated hadron 
resonance gas, $\chi_{nm}^{BC}$
with $n+m$ even and larger than 2 is approximately equal to $\chi_{11}^{BC}$. 
Thus three quantities, $\chi_{22}^{BC}/\chi_{13}^{BC}$, $\chi_{31}^{BS}/\chi_{11}^{BS}$ and $\chi_{31}^{BQ}/\chi_{11}^{BQ}$, which receive
contributions only from the charm, strange and (dominantly) light quark 
sectors, respectively, should be equal to unity as long as 
an uncorrelated hadron resonance gas model provides an appropriate description
of the thermodynamics of the medium. 
It is obvious from the right panel in Fig.~\ref{fig:decon} that
all three quantities start to deviate from unity
in the chiral crossover region.
This indicates that a description in terms of a HRG model breaks down for 
baryonic correlations involving light, strange, or charm quarks, i.e. open charm/strange hadrons start to 
dissociate in or just above the chiral crossover region.

\section{Thermodynamic contributions from unobserved hadrons near QCD transition}

A hadron resonance gas model that approximates QCD should include all states 
that are predicted by Quantum Chromodynamics. 
However, there are  quite a few states that are predicted in relativistic Quark 
Model (QM) and lattice QCD calculations \cite{Padmanath:2013bla} that have not
yet been observed in experiments~\cite{QM} and thus do not show up in the
particle data tables.
It thus is interesting to see whether 
these additional states can have any significant contribution to QCD 
thermodynamics~\cite{charmness,MissingS}.

\begin{figure}[!th]
\begin{center} 
\includegraphics[width=0.26\textwidth]{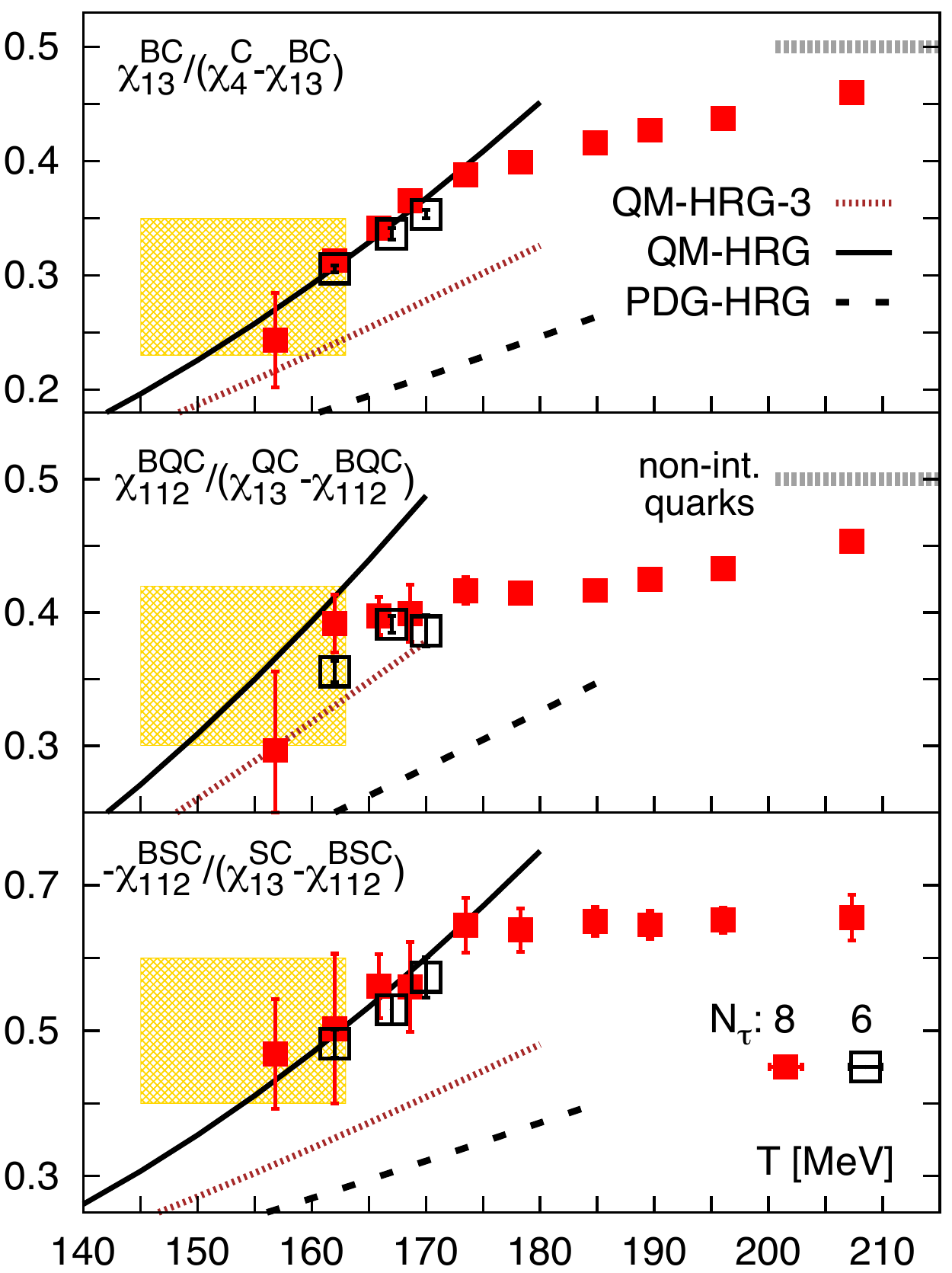}~\includegraphics[width=0.38\textwidth]{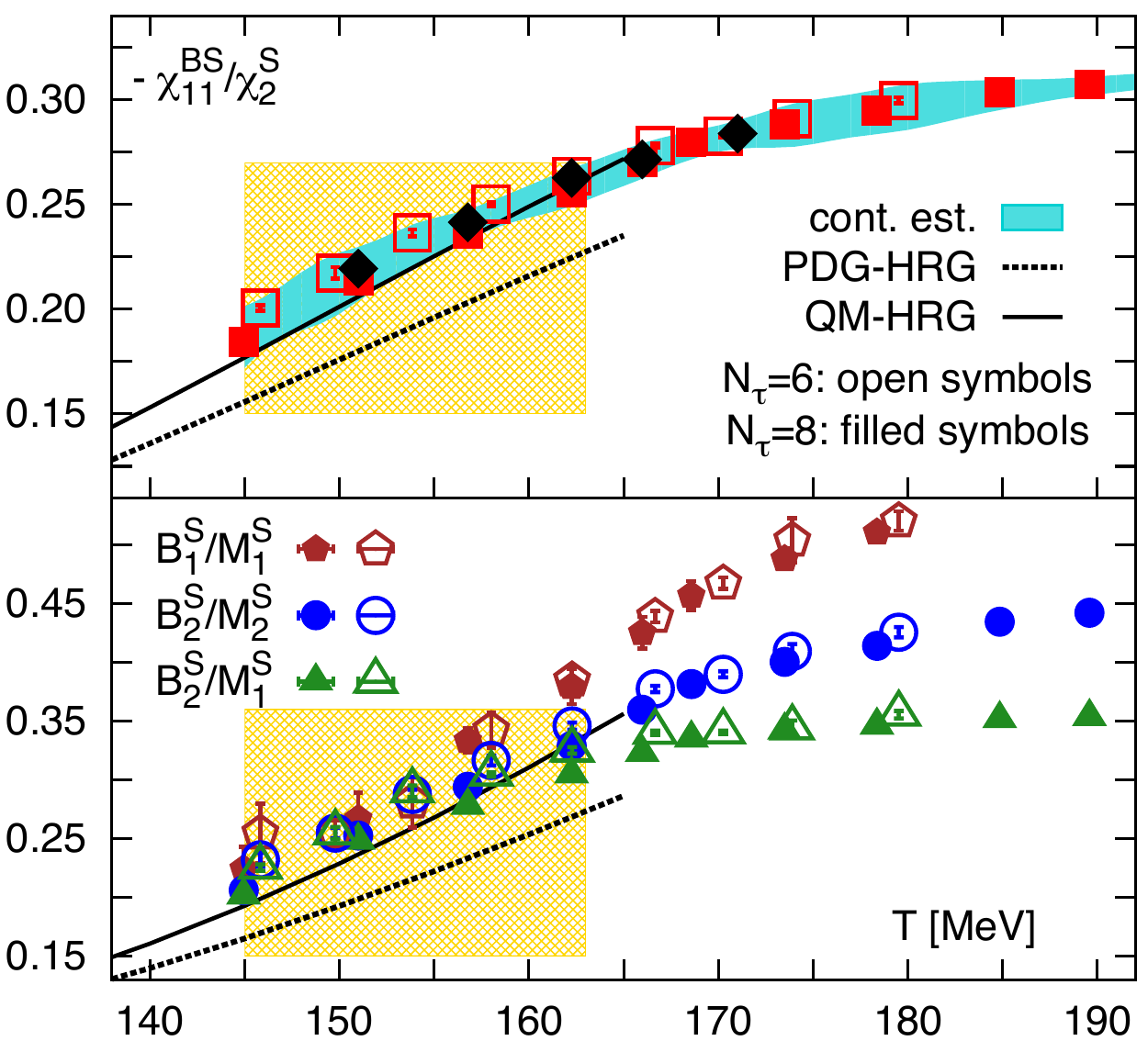}
\caption{Observation of thermodynamic contributions from open charm (left) and open strange (right) hadrons that are predicted in the relativistic Quark Model (QM) but not listed in the PDG table.
Figures are taken from Ref.~\cite{charmness} and Ref.~\cite{MissingS}, respectively. Yellow bands shown in these two plots represent the temperature 
window of the chiral crossover $T_c=154\pm 9$ MeV.}
\label{fig:abundance}
\end{center}
\end{figure}

One can check the partial pressures of open charm and strange hadrons in HRG models based on a particle spectrum predicted
in QM models (QM-HRG) and the spectrum listed in the PDG table (PDG-HRG). 
It turns out that there are only small difference in partial meson pressures 
while differences in baryon sector are much larger. 
This simply reflects the fact that the experimental knowledge of the 
strange and charm meson spectrum is more
complete than that for baryons. 
The difference is more pronounced in the ratio of partial pressures 
of baryons and mesons~\cite{charmness,MissingS}. In this spirit one can 
also construct observables that reflect the ratio of partial pressures of 
charmed baryons and mesons. 
For instance, quantities  like
$\chi_{13}^{BC}/(\chi_4^C-\chi_{13}^{BC})$, $\chi_{112}^{BQC}/(\chi_{13}^{QC}-\chi_{112}^{BQC})$ and $-\chi_{112}^{BSC}/(\chi_{13}^{SC}-\chi_{112}^{BSC})$
give the relative contributions of charmed baryons to open charm mesons, 
charged-charmed baryons to open charm charged mesons and 
strange-charmed baryons to 
strange-charmed mesons, respectively. As seen in the left panel of Fig.~\ref{fig:abundance} the temperature dependence of these three quantities
in the chiral crossover region can be better described by the solid line 
(QM-HRG), i.e. results obtained from an HRG model using the particle
spectrum predicted in the QM model. The popularly used HRG model based on the 
particle spectrum listed in the PDG table, i.e. PDG-HRG shown in the
plot, fails to describe the lattice data. The right panel in 
Fig.~\ref{fig:abundance} shows similar quantities for open strange 
hadrons: $-\chi_{11}^{BS}/\chi_2^S$ and 
$B_i^{S}/M_i^{S}$ reflect the relative contributions of strange baryons to open strange mesons in strange-baryon correlations and in partial pressures, respectively.
The same conclusion can be drawn from this plot as for the charm sector. 
The QM-HRG agrees better with the lattice data than the PDG-HRG. This
provides clear evidence for contributions from non-PDG listed hadrons to 
QCD thermodynamics and the transition from hadronic matter to the Quark 
Gluon Plasma. The importance of these additional states has also been pointed 
out in Ref.~\cite{Majumder:2010ik}. An important consequence of 
these states for the determination of freeze-out conditions for strange hadrons 
will be discussed in the next section.

\section{Freeze-out conditions}

It has been proposed in Ref.~\cite{TfLQCD} that the freeze-out/hadronization 
conditions can be determined by comparing fluctuations of net conserved charges
calculated on the lattice to the corresponding experimental observables. 
For instance the mean value, variance and skewness of net electric charge 
can be directly related to cumulants of net electric charge fluctuations
which are just the derivatives of pressure with respect to charge chemical 
potentials. They can be calculated on the lattice in 
a straightforward way. These quantities can also be measured in heavy ion 
experiments with a certain precision.

\begin{figure}[!th]
\begin{center}
\includegraphics[width=0.25\textwidth]{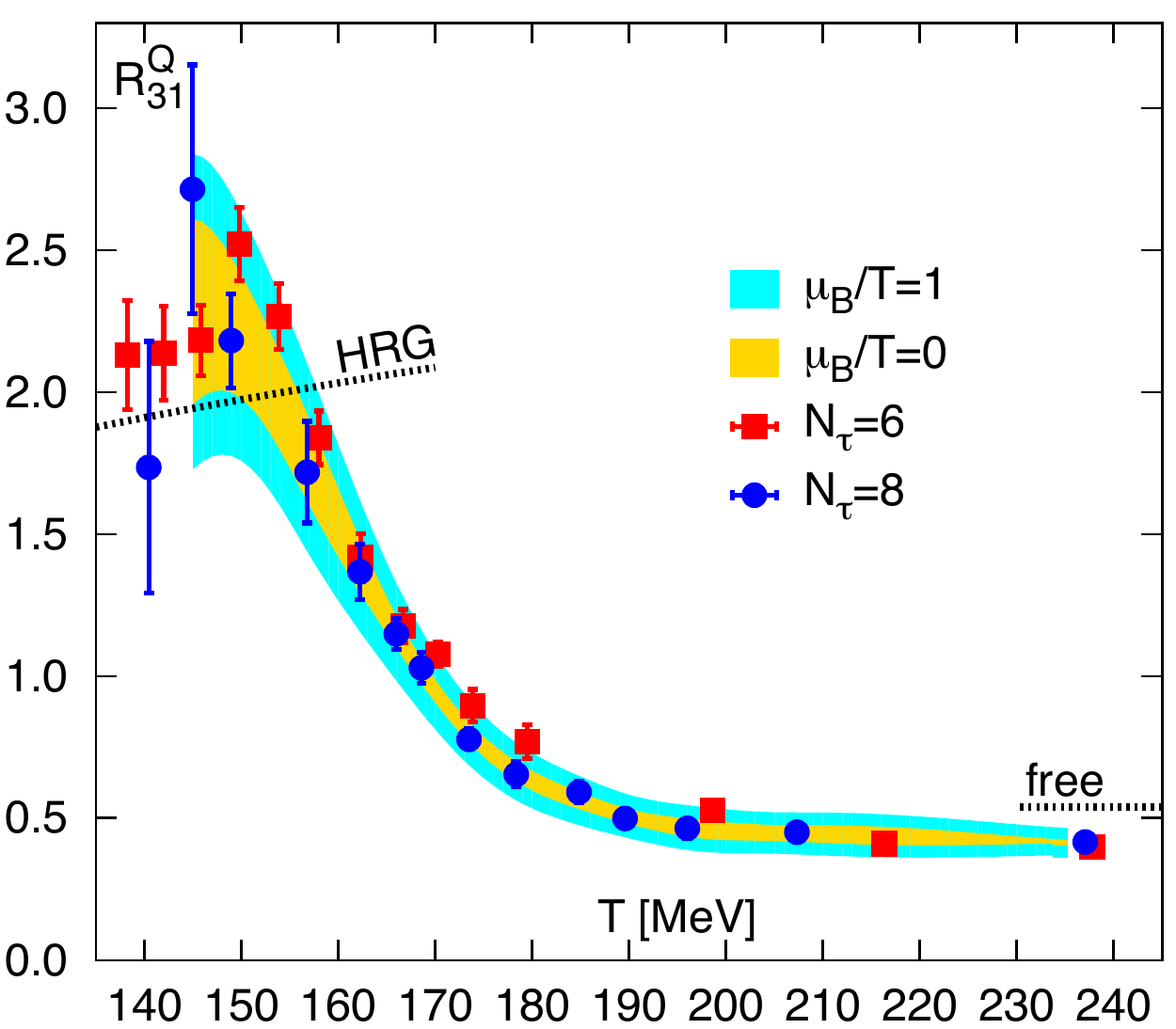}~
\includegraphics[width=0.25\textwidth]{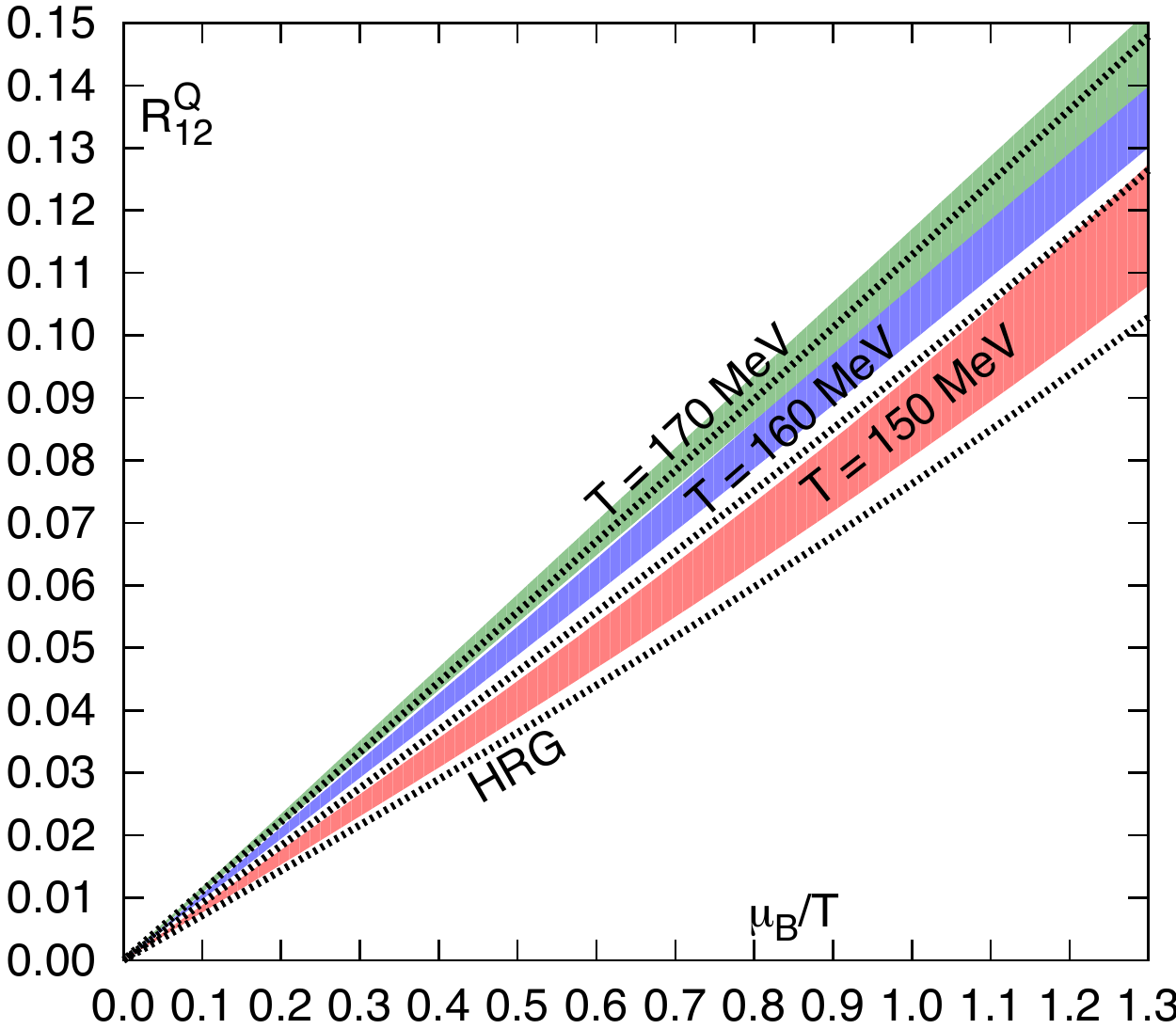}
~\includegraphics[width=0.33\textwidth]{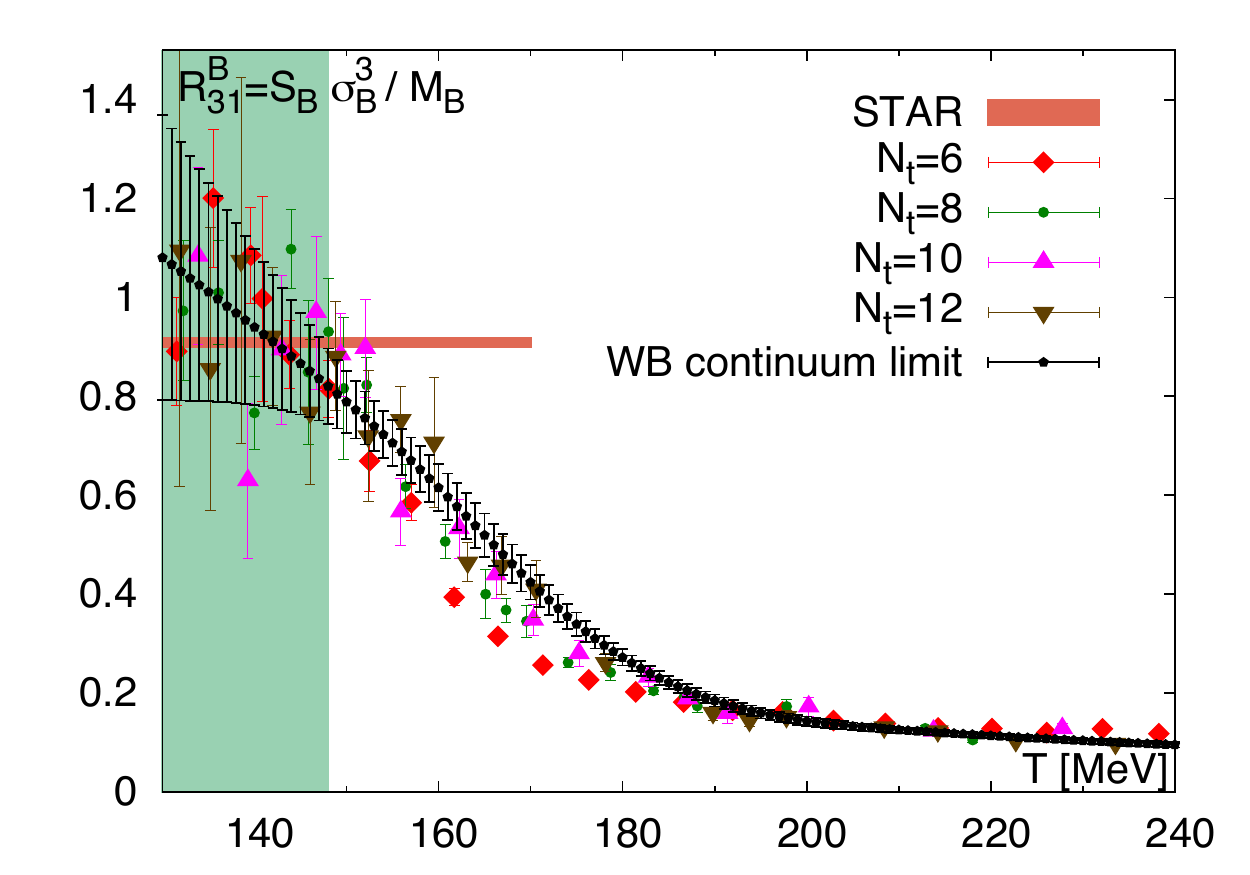}
\end{center}
\caption{The ratios $R_{31}^{Q}=S_Q\sigma^3_Q/M_Q$ versus temperature for $\mu_B=0$ (left) and $R_{12}^Q=M_Q/\sigma^2_Q$ versus $\mu_B/T$ for three values
of temperatures (middle) calculated on the lattice~\cite{TfLQCD}. Ratios $R_{31}^{Q}$ and  $R_{12}^{Q}$ can serve as thermometer to determine the temperature
and baryometer to determine the baryon chemical potential at the freeze-out, respectively. The right hand panel shows a comparison of lattice QCD data with the STAR measurements on the $S_B\sigma_B^3/M_B$ and the
upper limit of freeze-out temperature $T^{max}_f=148\pm4$ MeV is obtained from the comparison~\cite{Borsanyi:2014ewa}.}
\label{fig:freezeout}
\end{figure}

The ratio $R_{31}^Q=S_Q\sigma^3_Q/M_Q$ is related to skewness $S_Q$, variance  
$\sigma_Q$ and mean value $M_Q$ of net electric charge distributions
that are accessible in experimental measurements.  The left panel 
in Fig.~\ref{fig:freezeout} shows that $R_{31}^Q$ has only a weak dependence 
on the baryon 
density but changes rapidly with temperature. It thus can be used as a 
thermometer to extract the freeze-out temperatures of hadrons. 
On the other hand, the observable $R^Q_{12}=M_Q/\sigma^2_Q$ has a strong 
dependence on $\mu_B$ and it can be used as a baryometer to extract the 
baryon chemical potential at freeze-out. 
The approach to the determination of freeze-out conditions has also been 
pursued by using a different disrectization scheme on 
the lattice later on~\cite{Borsanyi:2013hza}.
However, current experimental data for net electrical charge fluctuations have
large errors and at present do not allow to perform a detailed comparison with 
lattice QCD data~\cite{Mukherjee:2013lsa,Adamczyk:2014fia}. 
To arrive at a better comparison of net charge fluctuation higher 
precision data  are certainly needed from e.g. BES-II~\cite{BESII}.
More accurate data are available on net proton number fluctuations~\cite{Adamczyk:2014ipa}. In the right panel of Fig.~\ref{fig:freezeout}
a comparison of lattice calculations of
net baryon number fluctuations and experimental measurements of net 
proton number fluctuations is shown. This provides an upper bound on the 
freeze-out temperature, i.e.  $T_f\leq148\pm4$ MeV~\cite{Borsanyi:2014ewa}. 
However, one needs to keep in mind that the net baryon and net proton 
fluctuations are not the same~\cite{Kitazawa}. 

\begin{figure}[!th]
\begin{center} 
\includegraphics[width=0.36\textwidth]{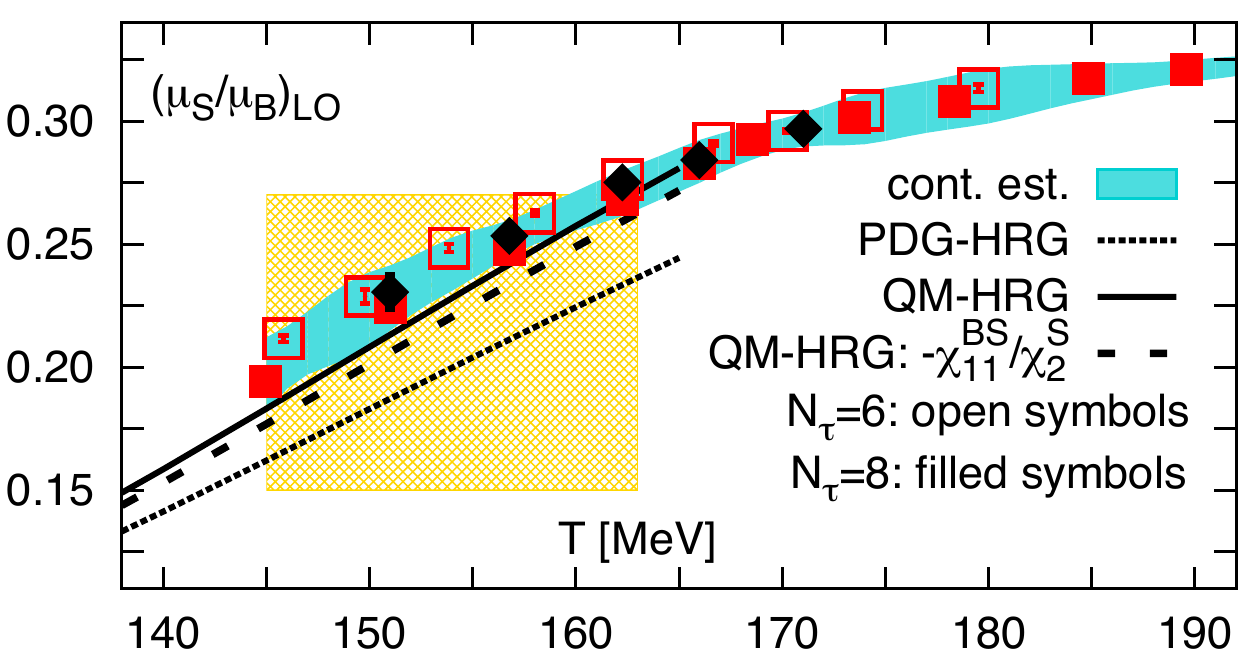}~\includegraphics[width=0.36\textwidth]{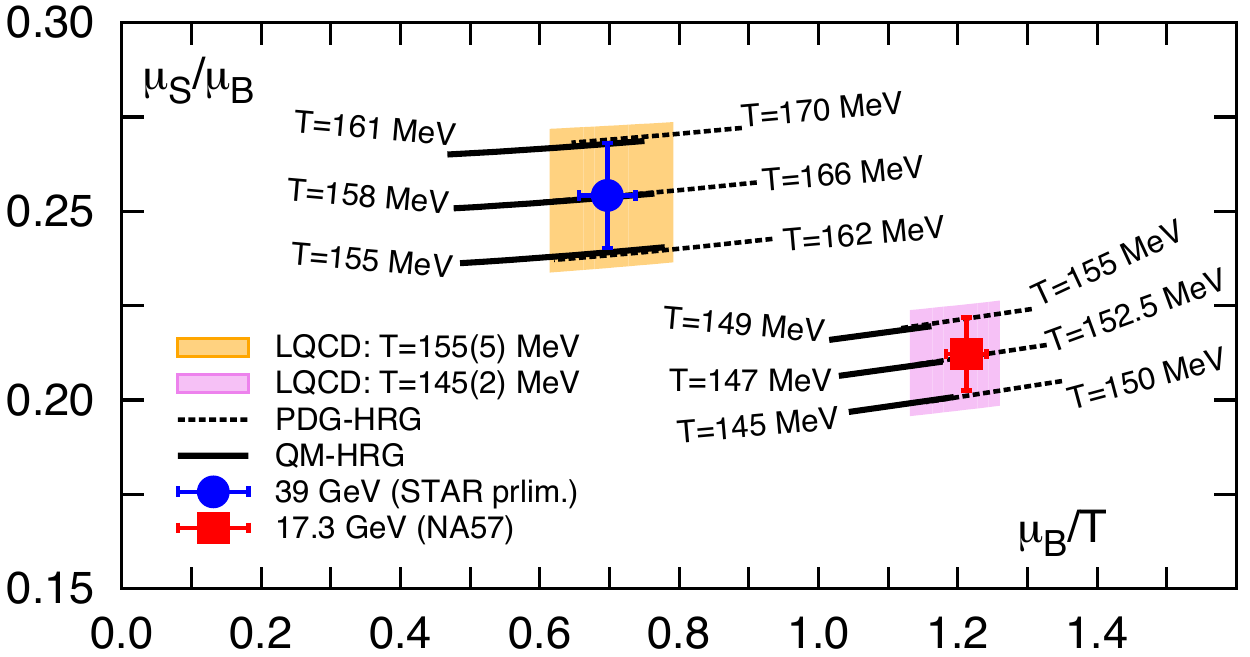}
\caption{Left: Imprints of contributions from unobserved strange hadrons to $-\chi_{11}^{BS}/\chi_2^S$ in $\mu_S/\mu_B$. Right: Comparison of
freeze-out temperature obtained from lattice QCD calculations (LQCD) and Hadron Resonance Gas model using the particle spectrum listed in 
the PDG table (PDG-HRG) and predicted from Quark Model (QM-HRG) by reproducing the relation of $\mu_S/\mu_B$  to $\mu_B/T$ determined from STAR and NA57
experiments. The temperatures obtained from QM-HRG are consistent with those from LQCD and are always smaller than those from PDG-HRG by around 5$\sim$8 MeV. Figures are taken from Ref.~\cite{strangeness}.}
\label{fig:Tf}
\end{center}
\end{figure}

One interesting question that has been discussed frequently in the recent 
literatures is 
whether strange hadrons freeze out at a higher temperature than light-quark
hadrons~\cite{Bellwied:2013cta,Alba:2014eba}. As shown in Fig.~\ref{fig:decon} and also in investigations of 
the screening masses of strange mesons~\cite{Mscr}, strange hadrons start to 
get deconfined in the same temperature 
window as light-quark hadrons do. This suggests that they will probably 
hadronize or freeze out at almost the same temperature. 
On the other hand, the indication of a higher freeze-out temperature for strange hadrons found in 
Ref.~\cite{Alba:2014eba} is mainly based on the 
quality of fits to the experimental data using the PDG-HRG model, i.e. 
fits become better when
different freeze-out temperatures are assumed or certain particle ratios are 
left out from the fits.

In this context it is important to understand systematic effects that may
arise from thermodynamic contributions of additional, unobserved strange 
hadrons. Strangeness neutrality in heavy ion collisions enforces a
dependence of the strangeness chemical potential $\mu_S$ on baryon chemical 
potential $\mu_B$ and temperature $T$.  To leading order $\mu_S$ can then be 
expanded in a Taylor series of $\mu_B$ as follows~\cite{strangeness}

\begin{equation}
\frac{\mu_S}{\mu_B} \simeq -\frac{\chi_{11}^{BS}}{\chi_{2}^S} - \frac{\chi_{11}^{QS}}{\chi_2^S}\frac{\mu_Q}{\mu_B} + \mathcal{O}(\mu_B^2).
\label{eq:muSomuB}
\end{equation}
In this equation the
term $-\chi_{11}^{BS}/\chi_2^S$ probes the relative abundance of strange 
baryons to open strange mesons. The sensitivity to the hadron spectrum is 
evident from the right panel of Fig.~\ref{fig:abundance}. At a given value of 
the temperature the value of $\mu_S/\mu_B$ therefore is influenced by 
thermodynamic contributions that arise from additional strange hadrons.
It can be clearly seen in the left panel of Fig.~\ref{fig:Tf} that the 
contribution from the term $-\chi_{11}^{BS}/\chi_2^S$ dominates in 
$\mu_S/\mu_B$ and that the calculation based on QM-HRG describes the lattice 
QCD results much better than a PDG-HRG calculation. Consequently, at a fixed 
value of $\mu_S/\mu_B$ a smaller value of temperature will be extracted when 
using lattice QCD and QM-HRG model calculations than by using PDG-HRG model 
calculations.

The relation of $\mu_S/\mu_B$ with $\mu_B/T$ can be obtained from a two-parameter fit to ratios of yields of strange anti-baryons to baryons  using the following ansatz motivated from the HRG model~\cite{strangeness}:
\begin{equation}
R_H \equiv \frac{\bar{H}_S}{H_S} = \exp \left[ -2(\mu_B^f/T^f) \times \left(1-(\mu_S^f/\mu_B^f)|S|\right) \right ],
\end{equation}
where $|S|$ is the absolute value of strangeness carried by the strange
baryons. Since details of hadron spectrum cancel in the ratio $R_H$, 
the ansatz is valid for both PDG-HRG and QM-HRG. It allows to extract two 
fitting parameters $\mu_B^f/T^f$ and $\mu_S^f/\mu_B^f$.
Thus the freeze-out temperature $T^f$ of strange hadrons can be obtained by adjusting the value of temperature in lattice QCD and HRG model 
calculations to reproduce values of $\mu^f_S/\mu^f_B$ and $\mu^f_B/T^f$ 
obtained from fits to experimental data. 
The right panel of Fig.~\ref{fig:Tf} shows the reproduction of
the value of $\mu_S^f/\mu_B^f$ and $\mu_B^f/T^f$ by adjusting the value of temperature in the lattice QCD calculations and PDG-HRG as well as QM-HRG model
calculations. The resulting value of $T^f$ from QM-HRG model calculations is consistent with that from lattice QCD calculations, however, it is smaller than that from 
PDG-HRG model calculations by around 5$-$8 MeV. 
This is mainly due to the fact that the presence of additional, experimentally 
still unobserved states, gets imprinted 
in $\mu_S^f/\mu_B^f$ as shown in Eq.~(\ref{eq:muSomuB}) and the left panel of 
Fig.~\ref{fig:Tf}. The freeze-out temperature $T^f$ determined for strange 
hadrons is thus is smaller by about 5$-$8 MeV when experimentally yet 
unobserved strange hadrons are taken into account. It becomes similar to
the freeze-out temperature of light-quark hadrons.
In other words there maybe no flavor hierarchy in the freeze-out conditions 
and light-quark and strange hadrons may freeze out in the same temperature 
region.

\begin{figure}[!th]
\begin{center} 
\includegraphics[width=0.37\textwidth]{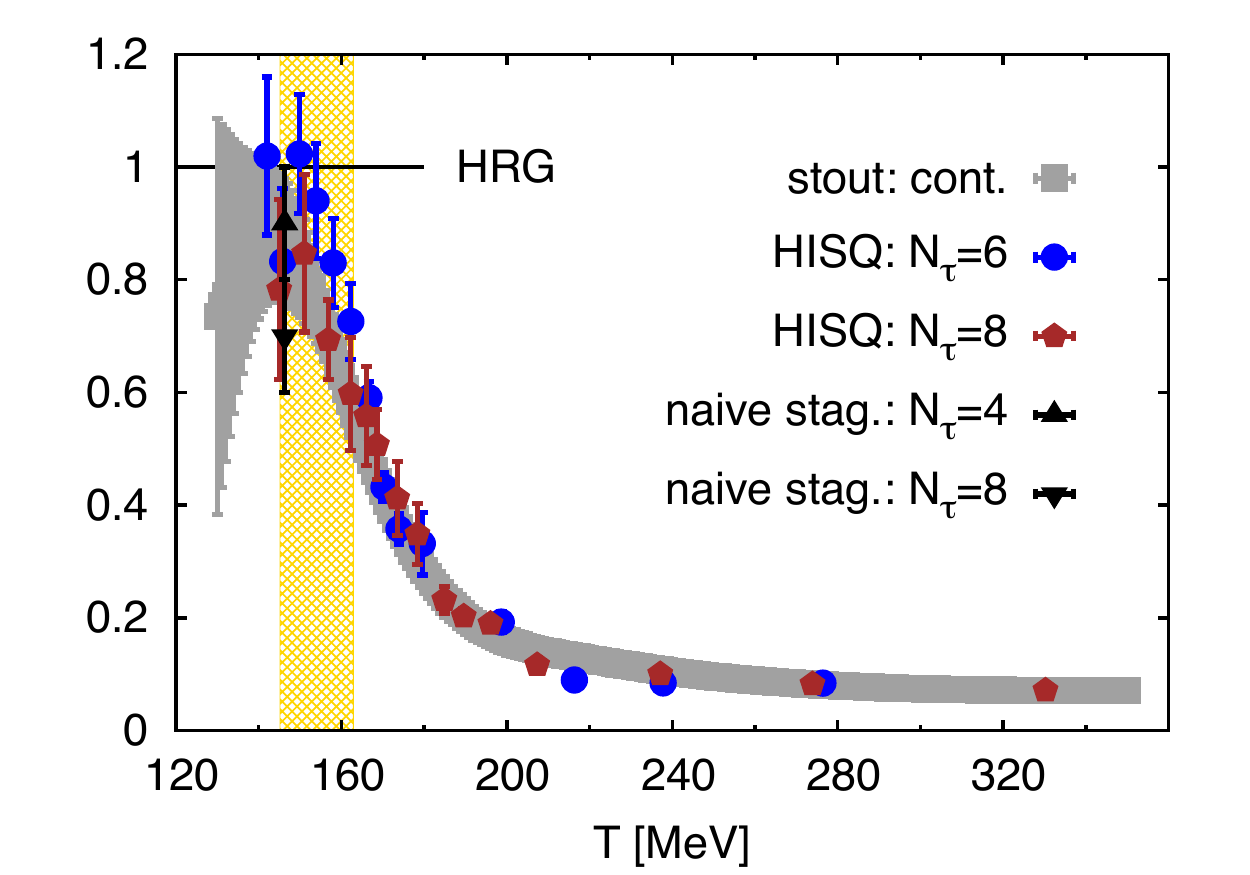}
\caption{$\chi^4_{B}/\chi^2_{B}$ as a function of temperature at $\mu_B=0$ obtained using the stout~\cite{Borsanyi:2013hza}, HISQ~\cite{strangeness} and na\"ive staggered~\cite{Gupta:2014qka} actions.}
\label{fig:fluctuations}
\end{center}
\end{figure}

\section{Towards the critical point}

To understand the dip shown in the experimental data of kurtosis of baryon 
number~\cite{Adamczyk:2014ipa} in terms of properties of conserved
charge fluctuations calculated on the lattice, one can expand kurtosis in a 
Taylor series of $\mu_B$ as follows
\begin{equation}
(\kappa\sigma^2)_B =\frac{\chi^B_{4,\mu}}{\chi^B_{2,\mu}} = \frac{\chi^B_4}{\chi^B_2} \left [  1 + \left(\frac{\chi^B_6}{\chi_4^B} -\frac{\chi^B_4}{\chi_2^B} \right) \left(\frac{\mu_B}{T}\right)^2 +  \cdots\right].
\end{equation}
The curvature of the kurtosis is then controlled by the difference of $\chi^B_6/\chi^B_4$ and $\chi^B_4/\chi^B_2$ up to the second order of $\mu_B/T$. 
A general agreement on the temperature dependence of $\chi^B_4/\chi_2^B$ is 
reached from lattice QCD calculations using different actions.
A determination of the 6th order cumulants still is missing and is crucially 
needed. Recent progress has also been made to directly simulate
the QCD phase diagram at nonzero baryon chemical potential using the 
Complex Langevin method~\cite{ComplexL}.

\section{Summary}

Since the last Quark Matter conference considerable progress has been made 
in understanding the equilibrium thermodynamics of strong interaction matter 
through lattice QCD calculations. Some highlights 
are: A consensus has been reached on the QCD Equation of State at vanishing 
baryon density 
by the HotQCD and Wuppertal-Budapest collaborations. The QCD EoS at nonzero 
baryon chemical potential, which
 is important for the BES program, has been calculated using Taylor expansions 
up to fourth order in $\mu_B/T$.
 The baryon number density can be reliably obtained in $\mathcal{O}(\mu_B^3)$ 
to describe heavy ion collisions down to $\sqrt{s_{NN}}\sim 30$~GeV;
The energy density at freeze-out remains roughly constant down 
to $\sqrt{s_{NN}}\sim50$ GeV. By looking at the correlations
 between charm or strangeness quantum numbers with baryon or electrical charge 
quantum numbers the onset of the melting of open charm and strange hadrons
can be identified. In both cases it happens in the chiral crossover 
temperature region. Evidence has been found for the thermodynamic relevance
of experimentally yet unobserved strange hadrons. They may influence the
determination of freeze-out conditions from experimentally measured particle 
yields.

%\section*{Acknowledgements}



%% References


%\bibliographystyle{elsarticle-num}

\begin{thebibliography}{00}

\bibitem{DWF_Tc}
  T.~Bhattacharya, M.~I.~Buchoff, N.~H.~Christ, H.~-T.~Ding, R.~Gupta, C.~Jung, F.~Karsch and Z.~Lin {\it et al.},
  %``The QCD phase transition with physical-mass, chiral quarks,''
  arXiv:1402.5175 [hep-lat].

\bibitem{axial}
M.~I.~Buchoff, M.~Cheng, N.~H.~Christ, H.~-T.~Ding, C.~Jung, F.~Karsch, Z.~Lin and R.~D.~Mawhinney {\it et al.},
  %``The QCD chiral transition, $\ua$ symmetry and the Dirac spectrum using domain wall fermions,''
  Phys.\ Rev.\ D {\bf 89} (2014) 054514,
  A.~Bazavov {\it et al.}  [HotQCD Collaboration],
  %``The chiral transition and $U(1)_A$ symmetry restoration from lattice QCD using Domain Wall Fermions,''
Phys.\ Rev.\ D {\bf 86} (2012) 094503,   G.~Cossu {\it et al.},
  %``Finite temperature study of the axial U(1) symmetry on the lattice with overlap fermion formulation,''
  Phys.\ Rev.\ D {\bf 87} (2013) 11,  114514.






  %%%%% EOS


\bibitem{firstlQCDEoS}
  J.~Engels, F.~Karsch, H.~Satz and I.~Montvay,
  %``High Temperature SU(2) Gluon Matter on the Lattice,''
  Phys.\ Lett.\ B {\bf 101} (1981) 89.

  
\bibitem{EOS_WB}
  S.~Borsanyi, Z.~Fodor, C.~Hoelbling, S.~D.~Katz, S.~Krieg and K.~K.~Szabo,
  %``Full result for the QCD equation of state with 2+1 flavors,''
  Phys.\ Lett.\ B {\bf 730} (2014) 99
  [arXiv:1309.5258 [hep-lat]].
    



\bibitem{EOS_hotQCD}
  A.~Bazavov, T.~Bhattacharya, C.~DeTar, H.~-T.~Ding, S.~Gottlieb, R.~Gupta, P.~Hegde and U.~M.~Heller {\it et al.},
  %``The equation of state in (2+1)-flavor QCD,''
  arXiv:1407.6387 [hep-lat].

  
  \bibitem{EOS_Bazavov}
A. Bazavov, this proceedings.

\bibitem{Cleymans:1999st}
  J.~Cleymans and K.~Redlich,
  %``Chemical and thermal freezeout parameters from 1-A/GeV to 200-A/GeV,''
  Phys.\ Rev.\ C {\bf 60} (1999) 054908.


\bibitem{EOS_fmu}
P. Hegde,  this proceedings.

\bibitem{EOS_fmuWB}
  S.~Borsanyi, G.~Endrodi, Z.~Fodor, S.~D.~Katz, S.~Krieg, C.~Ratti and K.~K.~Szabo,
  %``QCD equation of state at nonzero chemical potential: continuum results with physical quark masses at order $mu^2$,''
  JHEP {\bf 1208} (2012) 053.








  %%%%% Deconfinement
   
  
  
   
  \bibitem{WB_Tc} 
  S.~Borsanyi {\it et al.}  [Wuppertal-Budapest Collaboration],
  %``Is there still any T_c mystery in lattice QCD? Results with physical masses in the continuum limit III,''
  JHEP {\bf 1009}(2010) 073 . 
    
  \bibitem{HotQCD_Tc}
  A.~Bazavov, T.~Bhattacharya, M.~Cheng, C.~DeTar, H.~T.~Ding, S.~Gottlieb, R.~Gupta and P.~Hegde {\it et al.},
  %``The chiral and deconfinement aspects of the QCD transition,''
  Phys.\ Rev.\ D {\bf 85} (2012) 054503.
  


  


%%% quarkonia
  
  \bibitem{HQ_seed}
  T.~Matsui and H.~Satz,
  %``$J/\psi$ Suppression by Quark-Gluon Plasma Formation,''
  Phys.\ Lett.\ B {\bf 178} (1986) 416.
 
  
  
  \bibitem{Ding:2014xha}
  See e.g. H.~T.~Ding,
  %``Hard and thermal probes of QGP from the perspective of Lattice QCD,''
  arXiv:1404.5134 [hep-lat] and references therein.
  
  
    
\bibitem{Mscr} 
  Y.~Maezawa {\it et al.},
  %``Meson screening masses at finite temperature with Highly Improved Staggered Quarks,''
  arXiv:1312.4375,  F.~Karsch {\it el al.},
  %``Signatures of charmonium modification in spatial correlation functions,''
  Phys.\ Rev.\ D {\bf 85} (2012) 114501,
 % \bibitem{Cheng:2010fe}
  M.~Cheng {\it et al.},
  %``Meson screening masses from lattice QCD with two light and the strange quark,''
  Eur.\ Phys.\ J.\ C {\bf 71} (2011) 1564.


  
    \bibitem{Burnier:2012az}
  Y.~Burnier and A.~Rothkopf,
  %``Disentangling the timescales behind the non-perturbative heavy quark potential,''
  Phys.\ Rev.\ D {\bf 86} (2012) 051503
  [arXiv:1208.1899 [hep-ph]]. 
    
  
 \bibitem{Burnier:2013nla}
  Y.~Burnier and A.~Rothkopf,
  %``Bayesian Approach to Spectral Function Reconstruction for Euclidean Quantum Field Theories,''
  Phys.\ Rev.\ Lett.\  {\bf 111} (2013) 18,  182003
  [arXiv:1307.6106 [hep-lat]].
  

  

  
  
  \bibitem{charm}
  H.~-T.~Ding, A.~Francis, O.~Kaczmarek, F.~Karsch, H.~Satz and W.~Soeldner,
  %``Charmonium properties in hot quenched lattice QCD,''
  Phys.\ Rev.\ D {\bf 86} (2012) 014509,
    H.~Ohno,
  %``Quarkonium correlation functions at finite temperature in the charm to bottom region,''
  arXiv:1311.4565.
  %%CITATION = ARXIV:1311.4565;%%
  
  
    
  \bibitem{charm2}
  S.~Borsanyi, S.~Dürr, Z.~Fodor, C.~Hoelbling, S.~D.~Katz, S.~Krieg, S.~Mages and D.~Nogradi {\it et al.},
  %``Charmonium spectral functions from 2+1 flavour lattice QCD,''
  JHEP {\bf 1404} (2014) 132.
  

  
\bibitem{Bottomonia}
  G.~Aarts et al.,
  %``The bottomonium spectrum at finite temperature from $N_f=2+1$ lattice QCD,''
  arXiv:1402.6210 [hep-lat], 
  %G.~Aarts, C.~Allton, S.~Kim, M.~P.~Lombardo, S.~M.~Ryan and J.~-I.~Skullerud,
  %``Melting of P wave bottomonium states in the quark-gluon plasma from lattice NRQCD,''
  JHEP {\bf 1312} (2013) 064,  
  %G.~Aarts, C.~Allton, S.~Kim, M.~P.~Lombardo, M.~B.~Oktay, S.~M.~Ryan, D.~K.~Sinclair and J.~I.~Skullerud,
  %``What happens to the Upsilon and eta_b in the quark-gluon plasma? Bottomonium spectral functions from lattice QCD,''
  JHEP {\bf 1111} (2011) 103.  
  
 
%%% open hadrons  


  

    \bibitem{strangeness}
  A.~Bazavov, H.~-T.~Ding, P.~Hegde {\it et al.},
  %``Strangeness at high temperatures: from hadrons to quarks,''
  Phys.\  Rev.\  Lett.\  111, {\bf 082301} (2013).
  
  
  \bibitem{Bellwied:2013cta}
  R.~Bellwied, S.~Borsanyi, Z.~Fodor, S.~D.~Katz and C.~Ratti,
  %``Is there a flavor hierarchy in the deconfinement transition of QCD?,''
  Phys.\ Rev.\ Lett.\  {\bf 111} (2013) 202302.
  
  
    
  \bibitem{charmness}
  A.~Bazavov, H.~-T.~Ding, P.~Hegde {\it et al.},
  %``The melting and abundance of open charm hadrons,''
  arXiv:1404.4043 [hep-lat],   S.~Sharma,
  %``The thermodynamics of heavy light hadrons at freezeout,''
  arXiv:1408.2332 [hep-lat], this proceedings.
  
  
  

 
 %%%%%%%%%%%%%%%% section 4: unboserved hadrons
 
  \bibitem{Padmanath:2013bla}
  See e.g. M.~Padmanath, R.~G.~Edwards, N.~Mathur and M.~Peardon,
  %``Excited-state spectroscopy of singly, doubly and triply-charmed baryons from lattice QCD,''
  arXiv:1311.4806 [hep-lat].
   
 \bibitem{QM}
  See e.g. D.~Ebert, R.~N.~Faustov and V.~O.~Galkin,
  %``Heavy-light meson spectroscopy and Regge trajectories in the relativistic quark model,''
  Eur.\ Phys.\ J.\ C {\bf 66} (2010) 197,
  %[arXiv:0910.5612 [hep-ph]].  D.~Ebert, R.~N.~Faustov and V.~O.~Galkin,
  %``Spectroscopy and Regge trajectories of heavy baryons in the relativistic quark-diquark picture,''
  Phys.\ Rev.\ D {\bf 84} (2011) 014025.
% [arXiv:1105.0583 [hep-ph]].
 


   
  \bibitem{MissingS}
  A.~Bazavov, H.-T.~Ding, P.~Hegde {\it et al.},
  %``Additional Strange Hadrons from QCD Thermodynamics and Strangeness Freeze-out in Heavy Ion Collisions,''
  Phys.\ Rev.\ Lett.\ {\bf 113} (2014) 072001 [arXiv:1404.6511 [hep-lat]], 
  C. Schmidt, this proceedings.
 
 
 
 \bibitem{Majumder:2010ik}
  A.~Majumder and B.~Muller,
  %``Hadron Mass Spectrum from Lattice QCD,''
  Phys.\ Rev.\ Lett.\  {\bf 105} (2010) 252002,
  %\bibitem{Beitel:2014kza}
  M.~Beitel, K.~Gallmeister and C.~Greiner,
  %``Thermalization of Hadrons via Hagedorn States,''
  arXiv:1402.1458 [hep-ph].  
 
  
%%%%%%%%%%%%%%%% Section 5: Freeze out


  
  
  
  \bibitem{TfLQCD}
  A.~Bazavov, H.~T.~Ding, P.~Hegde {\it et al.},
  %``Freeze-out Conditions in Heavy Ion Collisions from QCD Thermodynamics,''
  Phys.\ Rev.\ Lett.\  {\bf 109} (2012) 192302.
    
  
  

  \bibitem{Borsanyi:2013hza}
  S.~Borsanyi, Z.~Fodor, S.~D.~Katz, S.~Krieg, C.~Ratti and K.~K.~Szabo,
  %``Freeze-out parameters: lattice meets experiment,''
  Phys.\ Rev.\ Lett.\  {\bf 111} (2013) 062005.


\bibitem{Mukherjee:2013lsa}
  S.~Mukherjee and M.~Wagner,
  %``Decon?nement of strangeness and freeze-out from charge ?uctuations,''
  PoS CPOD {\bf 2013} (2013) 039
  [arXiv:1307.6255 [nucl-th]].
 
 
\bibitem{Adamczyk:2014fia}
  L.~Adamczyk {\it et al.}  [STAR Collaboration],
  %``Beam energy dependence of moments of the net-charge multiplicity distributions in Au+Au collisions at RHIC,''
  arXiv:1402.1558 [nucl-ex]. 
 
 
 
\bibitem{BESII}
  G.~Odyniec,
  %``The RHIC Beam Energy Scan program in STAR and what's next ...,''
  J.\ Phys.\ Conf.\ Ser.\  {\bf 455} (2013) 012037.
  
  
   \bibitem{Adamczyk:2014ipa}
  L.~Adamczyk {\it et al.}  [STAR Collaboration],
  %``Beam-Energy Dependence of Directed Flow of Protons, Antiprotons and Pions in Au+Au Collisions,''
  Phys.\ Rev.\ Lett.\  {\bf 112} (2014) 162301


\bibitem{Borsanyi:2014ewa}
  S.~Borsanyi, Z.~Fodor, S.~D.~Katz, S.~Krieg, C.~Ratti and K.~K.~Szabo,
  %``Freeze-out parameters from electric charge and baryon number fluctuations: is there consistency?,''
  Phys.\ Rev.\ Lett.\  {\bf 113} (2014) 052301,   C. Ratti, this proceedings


 
  \bibitem{Kitazawa}
  M. Kitazawa, this proceedings.
  


  
  
  
  
   \bibitem{Alba:2014eba}
  P.~Alba {\it et al.}, 
  %``Freeze-out conditions from net-proton and net-charge fluctuations at RHIC,''
  arXiv:1403.4903 [hep-ph], %  \bibitem{Bugaev:2013sfa}
  K.~A.~Bugaev et al., 
  %D.~R.~Oliinychenko, J.~Cleymans, A.~I.~Ivanytskyi, I.~N.~Mishustin, E.~G.~Nikonov and V.~V.~Sagun,
  %``Chemical Freeze-out of Strange Particles and Possible Root of Strangeness Suppression,''
  Europhys.\ Lett.\  {\bf 104} (2013) 22002,
 % \bibitem{Chatterjee:2013yga}
  S.~Chatterjee, R.~M.~Godbole and S.~Gupta,
  %``Strange freezeout,''
  Phys.\ Lett.\ B {\bf 727} (2013) 554,
 %\bibitem{Andronic:2012dm}
  A.~Andronic, P.~Braun-Munzinger, K.~Redlich and J.~Stachel,
  %``The statistical model in Pb-Pb collisions at the LHC,''
  Nucl.\ Phys.\ A {\bf 904-905} (2013) 535c.
  %[arXiv:1210.7724 [nucl-th]].
  
  
  
%%%%%%%%%%%%%%%%%%%%%%%%% chi4B/chi2B  
 
  \bibitem{Gupta:2014qka}
  S.~Gupta, N.~Karthik and P.~Majumdar,
  %``The equation of state of QCD at finite chemical potential,''
  Phys.\ Rev.\ D {\bf 90} (2014) 034001
  [arXiv:1405.2206 [hep-lat]].
  
    \bibitem{ComplexL}
D. Sexty, this proceedings.

\end{thebibliography}



\end{document}